\documentstyle[a4p,epsfig]{article}

\begin{document}
%=======================================================================
%       Parameters affecting the general appearance
%=======================================================================
%\setlength{\parskip}{\medskipamount}

%=======================================================================
%       Define symbols
%=======================================================================

\newcommand{\roots}{\sqrt{\mathrm s}}
\newcommand{\qqbar}{{\mathrm q}\bar{\mathrm q}}
\newcommand{\ee}{{\mathrm e}^+{\mathrm e}^-}
\newcommand{\mumu}{\mu^+\mu^-}
\newcommand{\tautau}{\tau^+\tau^-}
\newcommand{\lplm}{l^+l^-}

\newcommand{\MZ}{{\mathrm M}_{\mathrm Z}}
\newcommand{\Zzero}{{\mathrm Z}^0}

\newcommand{\eetoee}{\ee\to\ee}
\newcommand{\eetoll}{\ee\to\lplm}
\newcommand{\eegammall}{\ee\to\gamma^*\to\lplm}
\newcommand{\eetomumu}{\ee\to\mumu}
\newcommand{\eetotautau}{\ee\to\tautau}
\newcommand{\eetoqqbar}{\ee\to\qqbar}

\newcommand{\tausevenp}{\mbox{$\tau^- \to 4\pi^-\;3\pi^+\;n\gamma\;\nu_\tau\;(n \geq 0)$}}
\newcommand{\tautosevenpi}{\mbox{$\tau^- \to 4\pi^-\;3\pi^+\;\nu_\tau$}}
\newcommand{\sevenpi}{\mbox{$4\pi^-\;3\pi^+\;\nu_\tau$}}
\newcommand{\sevenpipizero}{\mbox{$4\pi^-\;3\pi^+\;\pi^0\;\nu_\tau$}}
\newcommand{\tautosevenpipizero}{\mbox{$\tau^- \to 4\pi^-\;3\pi^+\;\pi^0\;\nu_\tau$}}
\newcommand{\taucleo}{\mbox{$\tau^- \to 3\pi^-\;2\pi^+\;2\pi^0\;\nu_\tau$}}

\newcommand{\tausevenpbr}{\mbox{{\it BR} (\tausevenp)}}
\newcommand{\taucleobr}{\mbox{{\it BR} (\taucleo )}}

\newcommand{\tausevenplimit}{1.8 \times 10^{-5}}
\newcommand{\ntauevents}{210680\pm1100}

\newcommand{\gev}{\mathrm GeV}
\newcommand{\gevc}{{\mathrm GeV}/c}
\newcommand{\gevcc}{{\mathrm GeV}/c^2}
\newcommand{\mevcc}{{\mathrm MeV}/c^2}
\newcommand{\ipb}{{\mathrm pb}^{-1}}

\newcommand{\rhad}{R_{\mathrm had}}
\newcommand{\fres}{f_{\mathrm res}}

\newcommand{\dedx}{{\mathrm d}E/{\mathrm d}x}
\newcommand{\dedxsstwo}{f_2}
\newcommand{\ntrkone}{N_1}
\newcommand{\ntrktwo}{N_2}
\newcommand{\necal}{N_{\mathrm ECAL}}
\newcommand{\dangmxone}{{\alpha}_1^{\mathrm max}}
\newcommand{\dangmxtwo}{{\alpha}_2^{\mathrm max}}

\newcommand{\qconeone}{|\sum_{\mathrm tracks} q |_1}
\newcommand{\qconetwo}{|\sum_{\mathrm tracks} q |_2}

\newcommand{\thetacone}{\theta_{\mathrm cone}}
\newcommand{\econe}{E_{\mathrm cone}}
\newcommand{\pcone}{P_{\mathrm cone}}
\newcommand{\eoptrktwo}{(\econe/\pcone)_2}

\newcommand{\invmastwo}{m_2}
\newcommand{\mtau}{m_{\tau}}
\newcommand{\mpi}{m_{\pi}}
\newcommand{\ntautau}{N_{\tau^+\tau^-}}

% miscellaneous
% Def. fuer groesser-ungefaehr:
\newcommand{\gsim}{\;\raisebox{-0.9ex}
           {$\textstyle\stackrel{\textstyle >}{\sim}$}\;}
% Def. fuer kleiner-ungefaehr:
\newcommand{\lsim}{\;\raisebox{-0.9ex}{$\textstyle\stackrel{\textstyle<}
           {\sim}$}\;}

\newcommand{\raiseb}{\raisebox{1.5ex}[-1.5ex]}
\def\pz{\phantom{0}}
\def\pzz{\phantom{00}}
\def\pzzz{\phantom{000}}
\def\ppo{\phantom{.}}

\newcommand{\degree}    {^\circ}
\newcommand{\Evis}      {E_{\mathrm{vis}}}
\newcommand{\Rvis}      {E_{\mathrm{vis}}\,/\roots}
\newcommand{\Mvis}      {M_{\mathrm{vis}}}
\newcommand{\Rbal}      {R_{\mathrm{bal}}}
\newcommand{\pt}       {p_{\mathrm{t}}}

%----------------------------
%  Bibliographic references
%----------------------------
%
%     Journal names
%
\newcommand{\PhysLett}  {Phys.~Lett.}
\newcommand{\PRL} {Phys.~Rev.~Lett.}
\newcommand{\PhysRep}   {Phys.~Rep.}
\newcommand{\PhysRev}   {Phys.~Rev.}
\newcommand{\NPhys}  {Nucl.~Phys.}
\newcommand{\NIM} {Nucl.~Instr.~Meth.}
\newcommand{\CPC} {Comp.~Phys.~Comm.}
\newcommand{\ZPhys}  {Z.~Phys.}
\newcommand{\etal}     {et al.}
%
%     Collaboration names
%
\newcommand{\OPALColl}  {OPAL Collaboration}
\newcommand{\ALEPHColl} {ALEPH Collaboration}
\newcommand{\DELPHIColl} {DELPHI Collaboration}
\newcommand{\LColl}     {L3 Collaboration}
\newcommand{\JADEColl}  {JADE Collaboration}
\newcommand{\HRSColl}  {HRS Collaboration}
\newcommand{\CLEOColl}  {CLEO Collaboration}

%=======================================================================
%       Title Page
%=======================================================================

\renewcommand{\Huge}{\huge}
\parskip12pt plus 1pt minus 1pt
\topsep0pt plus 1pt
\begin{titlepage}
\begin{center}{\large   EUROPEAN LABORATORY FOR PARTICLE PHYSICS
}\end{center}\bigskip
\begin{flushright}
       CERN-PPE/97-049   \\ 2nd May 1997
\end{flushright}
\bigskip\bigskip\bigskip\bigskip\bigskip
\begin{center}{\huge\bf\boldmath
        An upper limit on the
        branching ratio for $\tau$ decays
        into seven charged particles
}\end{center}\bigskip\bigskip
\begin{center}{\LARGE The OPAL Collaboration
}\end{center}\bigskip\bigskip
\bigskip\begin{center}{\large  Abstract}\end{center}
  We have searched for decays of the $\tau$ lepton into seven or more
  charged particles, using data collected with the OPAL detector from
  1990 to 1995 in $\ee$ collisions at $\roots~\approx~\MZ$.  No
  candidate events were found and an upper limit on the branching
  ratio for $\tau$ decays into seven charged particles of
  $\tausevenplimit$ at the 95\% confidence level was determined.
\bigskip\bigskip\bigskip\bigskip
\bigskip\bigskip
\begin{center}{\large
(Submitted to Physics Letters B)
}\end{center}
\end{titlepage}
\begin{center}{\Large        The OPAL Collaboration
}\end{center}\bigskip
\begin{center}{
%begin authorlist
K.\thinspace Ackerstaff$^{  8}$,
G.\thinspace Alexander$^{ 23}$,
J.\thinspace Allison$^{ 16}$,
N.\thinspace Altekamp$^{  5}$,
K.J.\thinspace Anderson$^{  9}$,
S.\thinspace Anderson$^{ 12}$,
S.\thinspace Arcelli$^{  2}$,
S.\thinspace Asai$^{ 24}$,
D.\thinspace Axen$^{ 29}$,
G.\thinspace Azuelos$^{ 18,  a}$
% ...
%end authorlist
}\end{center}\bigskip
\bigskip
%begin institutes
$^{  1}$School of Physics and Space Research, University of Birmingham,
Birmingham B15 2TT, UK
\newline
$^{  2}$Dipartimento di Fisica dell' Universit\`a di Bologna and INFN,
I-40126 Bologna, Italy
\newline
$^{  3}$Physikalisches Institut, Universit\"at Bonn,
D-53115 Bonn, Germany
\newline
$^{  4}$Department of Physics, University of California,
Riverside CA 92521, USA
\newline
$^{  5}$Cavendish Laboratory, Cambridge CB3 0HE, UK
\newline
% ...
$^{  8}$CERN, European Organisation for Particle Physics,
CH-1211 Geneva 23, Switzerland
\newline
$^{  9}$Enrico Fermi Institute and Department of Physics,
University of Chicago, Chicago IL 60637, USA
\newline
% ...
$^{ 12}$Indiana University, Department of Physics,
Swain Hall West 117, Bloomington IN 47405, USA
\newline
% ...
$^{ 16}$Department of Physics, Schuster Laboratory, The University,
Manchester M13 9PL, UK
\newline
% ...
$^{ 18}$Laboratoire de Physique Nucl\'eaire, Universit\'e de Montr\'eal,
Montr\'eal, Quebec H3C 3J7, Canada
\newline
% ...
$^{ 23}$Department of Physics and Astronomy, Tel Aviv University,
Tel Aviv 69978, Israel
\newline
$^{ 24}$International Centre for Elementary Particle Physics and
Department of Physics, University of Tokyo, Tokyo 113, and
Kobe University, Kobe 657, Japan
\newline
% ...
$^{ 29}$University of British Columbia, Department of Physics,
Vancouver BC V6T 1Z1, Canada
\newline
% ...
$^{  a}$ and at TRIUMF, Vancouver, Canada V6T 2A3
\newline
% ...
\newline
%end notes

%-----------------------------------------------------------------------
\clearpage
%=======================================================================
\section{Introduction}
%=======================================================================
\label{introduction}

During the years 1990 to 1995 of LEP operation, an integrated
luminosity of 163~$\ipb$ has been recorded with the OPAL detector at
$\roots~\approx 91~\gev$, corresponding to approximately five million
observed $\Zzero$ decays.  This large data sample, which contains more
than 210000 $\tau$ pair events, allows searches for rare decays of the
$\tau$ lepton with branching ratios down to about $10^{-5}$. The
high-multiplicity decays of the $\tau$ are particularly interesting
for a measurement of the $\tau$ neutrino mass.  Recent examples of
such measurements using $\tau$ decays into three or five charged
particles can be found in references~\cite{nutau3prong,nutau5prong}.
Taking only kinematical considerations into account, the decay of the
$\tau$ into twelve pions is possible.

In this note we present a search for $\tau$ decays into seven charged
particles (7-prong).  The HRS Collaboration has published an upper
limit of $1.9 \times 10^{-4}$ on the branching ratio for $\tau$ decays
into seven charged particles\footnote{References in this paper to specific
charge states apply to the charge conjugate states as well.},
\tausevenpbr, at the 90\% confidence level~\cite{hrs}.  The CLEO
Collaboration has determined a 95\% confidence level upper limit of
$1.1 \times 10^{-4}$ on the branching ratio
\taucleobr~\cite{cleo}.

%=======================================================================
\section{The OPAL detector}
%=======================================================================
\label{opaldet}

A detailed description of the OPAL detector can be found
elsewhere~\cite{opaldetector}. Sub-detectors which are particularly
relevant to the present analysis are described here briefly.  The
central detector consists of a system of tracking chambers providing
charged particle tracking over 96\% of the full solid angle inside a
0.435~T uniform magnetic field parallel to the beam axis\footnote{The
OPAL coordinate system is a right-handed coordinate system which is
defined so that the $z$-axis is in the direction of the electron beam,
the $x$-axis is horizontal and points towards the centre of the LEP
ring; $\theta$ and $\phi$ are the polar and azimuthal angles, defined
relative to the $+z$- and $+x$-axes, respectively.}.  Starting with the
innermost components, it consists of a high precision silicon
microvertex detector which was available from 1991, a precision vertex
drift chamber, a large volume jet chamber with 159 layers of axial
anode wires and a set of $z$ chambers measuring the track coordinates
along the beam direction.  The efficiency for separating hits from two
different particles in the jet chamber is approximately 80\% for
distances between two hits of 2.5~mm~\cite{cjperf} and drops rapidly
for smaller hit distances.  The jet chamber also provides energy loss
measurements which are used for particle identification.

A lead-glass electromagnetic calorimeter (ECAL) located outside the
magnet coil covers the full azimuthal range with excellent hermeticity
in the polar angle range of $|\cos \theta |<0.82$ for the barrel
region and $0.81<|\cos \theta |<0.984$ for the endcap region.  The
forward detectors and silicon tungsten calorimeters located at both
sides of the interaction point measure the luminosity and complete the
geometrical acceptance down to 24~mrad in polar angle.

%=======================================================================
\section{Event simulation}
%=======================================================================
\label{evsimulation}

The $\eetotautau$ Monte Carlo events were generated using
KORALZ~4.0~\cite{koralz}. The dynamics of $\tau$ decays to final
states with five or fewer charged particles were simulated with the
TAUOLA~2.4 decay library~\cite{tauola}. The 7-prong signal events were
simulated by modifying the TAUOLA routine describing the decay of the
$\tau$ into five or more pions to allow up to seven charged pions with
either zero (\tautosevenpi) or one neutral pions (\tautosevenpipizero)
in the final state.  To determine the efficiency for reconstructing an
event with a 7-prong $\tau$ decay we generated $\tau$ pair events in
which the first $\tau$ decays according to the standard TAUOLA library
and the second one decays into seven charged particles.

We have implemented various phase space models for these 7-prong
decays. The standard case assumes an isotropic phase space between the
minimum allowed value, equivalent to the sum of the pion masses, and
the maximum allowed value, $\mtau$. To study the effect of the phase
space modelling on the final result, we also considered the two
extreme cases of a ``low'' and ``high'' phase space, that is with a
7-pion invariant mass always below or above the value of $1.4~\gevcc$,
which is the mean value of $7\mpi$ and $\mtau$. Additionally, we
generated 7-prong Monte Carlo events where the invariant mass of the
7-pion system is peaked close to this mean value in a Gaussian manner
with a standard deviation of $150~\mevcc$.

To estimate the background from $\eetotautau$ events with $\tau$
decays into up to five charged particles, we have used a sample of
920000 events generated with KORALZ and TAUOLA.  The expected rate of
multihadronic background was determined using $\eetoqqbar$ events
generated with JETSET~\cite{jetset}. The background coming from
$\eetoee$ events was estimated using a sample of simulated $\eetoee$
events corresponding to a luminosity of 287~$\ipb$ and generated with
BABAMC~\cite{babamc}.  Contributions from two-photon processes were
studied using the PYTHIA~\cite{jetset} and
Vermaseren~\cite{vermaseren} Monte Carlo generators.  Background
contributions coming from $\eetomumu$ and four-fermion events are
small. They were investigated using the KORALZ and
FERMISV~\cite{fermisv} event generators, respectively.

All Monte Carlo events were passed through the GEANT~\cite{geant}
simulation of the OPAL detector~\cite{gopal} and were then processed
in the same way as the real data.

%=======================================================================
\section{Event preselection}
%=======================================================================
\label{preselection}

A cone jet algorithm~\cite{conealgo} was employed to assign all tracks
and electromagnetic clusters to cones with a half opening angle of
$35\degree$.  We required that exactly two cones with at least one
track each were found in the event.  Tracks were required to satisfy
the following conditions:
\begin{itemize}
\item $p_\perp > 0.1~\gevc$, where $p_\perp$ is the momentum 
component transverse to the beam direction.
\item A minimum number of 20 hits in the central jet chamber
was required.  This restricts the acceptance of the detector for tracks
to $|\cos\theta|~<~0.965$, where $\theta$ is the angle between the
track and the beam axis.
\item The distance of closest approach of the track to the 
beam axis must be less than 2~cm and the displacement of 
the track along the beam axis from the nominal interaction 
point at the point of closest approach to the beam must be
less than 20~cm. 
\end{itemize}

To reject events coming from two-photon interactions, the
acollinearity of the two cones was required to be less than
$15\degree$, with the direction of each cone given by the momentum sum
of all tracks and electromagnetic clusters inside the cone.  The
momenta of the ECAL clusters were calculated using the energy
deposition and the angles of the cluster.  The visible energy of each
jet is taken as the maximum of the sum of the track momenta and the
sum of the ECAL cluster energies associated with that jet.  Events were
rejected if the sum of the visible energies of the two jets was less
than 3\% of the centre-of-mass energy. Events with a total visible
energy smaller than 20\% of the centre-of-mass energy were rejected if
the sum of the transverse momenta of the tracks and the sum of the
transverse momenta of the electromagnetic clusters were both smaller
than $2~\gev$.  This set of cuts rejects most of the two-photon
background.  Cosmic ray events were removed using the time-of-flight
and the tracking chamber information~\cite{taulepton}.  Finally, we
rejected events where some particles may be lost close to the beam
pipe by requiring that the total energy deposit in the forward
detectors be less than $2~\gev$.  We also required
$|\cos\thetacone|~<~0.9$ for each of the two cones in the event, where
$\thetacone$ is the angle between the jet direction and the beam
axis. This cut ensures that both jets are located inside the
acceptance of the detector.

Only events with a minimum of four tracks were accepted for further
analysis.  The efficiency for selecting a $\tautau$ event with a
7-prong $\tau$ decay after the preselection cuts was $(80.8\pm0.5)\%$
according to the Monte Carlo simulation.

%=======================================================================
\section{Selection of 7-prong {\boldmath $\tau$} events}
%=======================================================================
\label{selection}

\begin{figure}[htb]
\centerline{\epsfig{figure=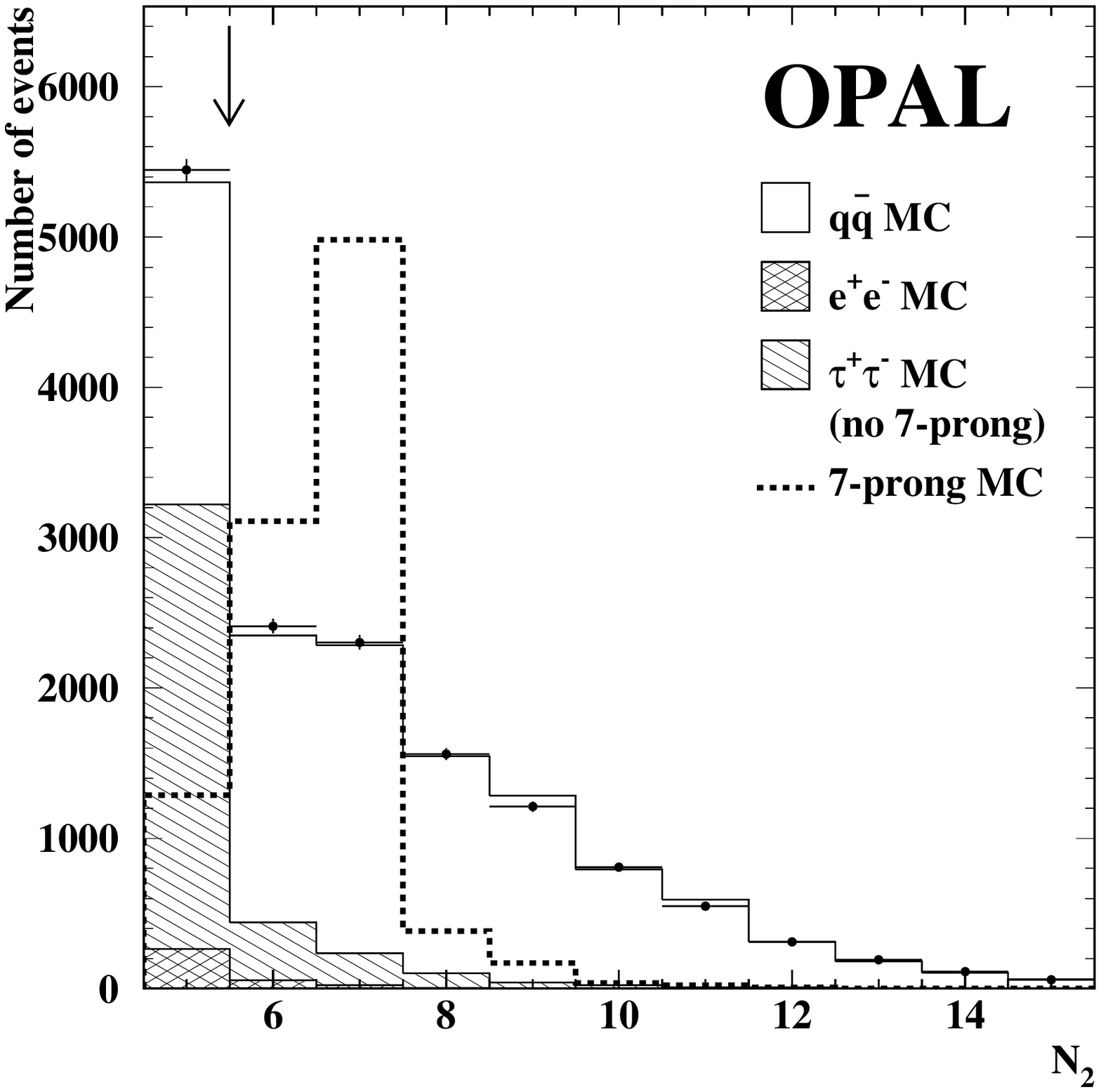,width=0.9\textwidth}} 
\caption{\it Distribution of the number of tracks, $\ntrktwo$, 
in the second cone after relaxing the cut to $\ntrktwo \geq 5$.  An
arbitrary normalisation is used for the 7-prong signal Monte Carlo
simulation. The other histograms show the contributions from the
various background Monte Carlo samples. The points are the data. The
arrow indicates the position of the selection cut.}
\label{fig:one} 
\end{figure}

At this stage, the largest sources of background, around 55\%, come
from $\eetoqqbar$ events with a low-multiplicity jet, followed by
$\eetotautau$ events with $\tau$ decays into five or fewer charged
particles ($\approx 33\%$).  These backgrounds were reduced by the
following set of cuts.  Since approximately 99.9\% of the $\tau$
leptons decay into one or three charged particles, we required that
the number of tracks $\ntrkone$ in one of the two cones\footnote{In
the rest of the paper we refer to this as the first cone.} is exactly
1 or 3.  The absolute value of the sum of the charges of these tracks
was required to satisfy $\qconeone = 1$.  The search for the 7-prong
$\tau$ decay was performed in the second cone. We required that this
contains at least six tracks ($\ntrktwo \geq 6$) and that the absolute
value of the sum of the track charges satisfies $\qconetwo \leq 2$.
The distribution of $\ntrktwo$ is shown in Figure~\ref{fig:one}.

\begin{figure}[htb]
\centerline{\epsfig{figure=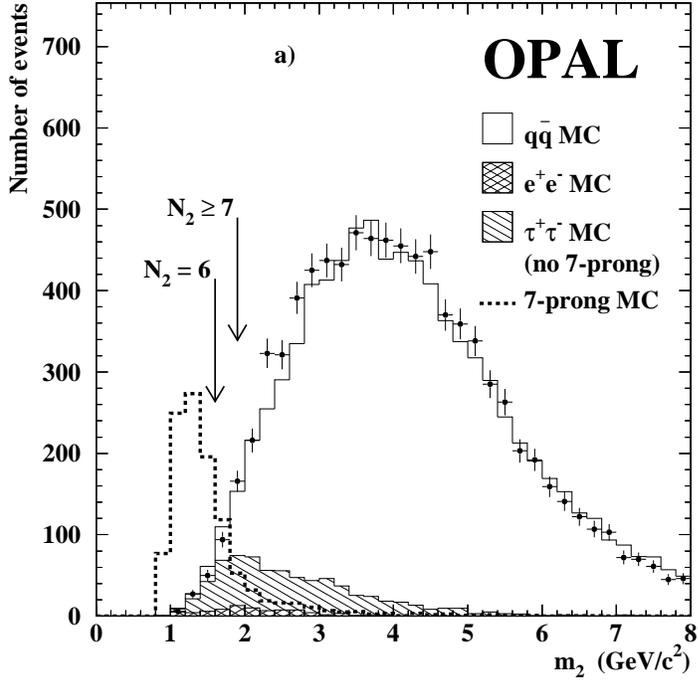,width=0.6\textwidth}}
\centerline{\epsfig{figure=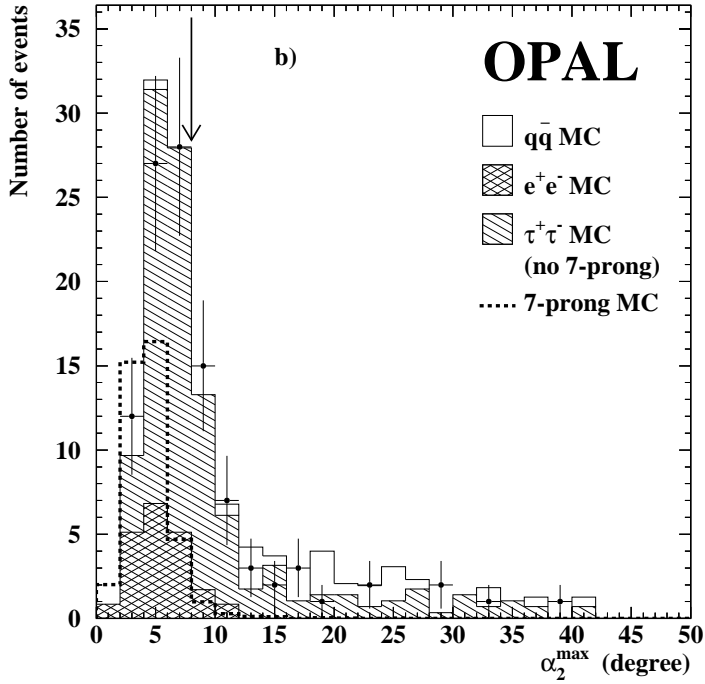,width=0.6\textwidth}}
\caption{\it 
Distribution of $\invmastwo$ 
before the invariant mass cut (a) and of $\dangmxtwo$ after the cuts
on $\invmastwo$ and $\dangmxone$ (b). An arbitrary normalisation is
used for the 7-prong signal Monte Carlo simulation.  The other
histograms show the contributions from the various background Monte
Carlo samples. The points are the data.  The arrows indicate the
positions of the selection cuts.}
\label{fig:two} 
\end{figure}

The invariant mass of the tracks from a true 7-prong $\tau$ decay must
be less than the $\tau$ mass. Therefore the invariant mass of the
tracks in the second cone, $\invmastwo$, calculated using the pion
mass for each track, was required to satisfy
$\invmastwo~\leq~1.6~\gevcc$ for cones with six reconstructed tracks
and $\invmastwo~\leq~1.9~\gevcc$ for cones with $\ntrktwo~\geq~7$.
The cut is lower for the $\ntrktwo=6$ case since the measured
invariant mass of a true 7-prong decay candidate would be smaller due
to the missing track.  The distribution of $\invmastwo$ before the cut
is shown in Figure~\ref{fig:two}.

Most of the $\qqbar$ events that were not removed by the invariant
mass cut were rejected by a cut on the maximum opening angle of any
pair of tracks in the two jets.  For the first cone we required this
angle to be less than $10\degree$ ($\dangmxone~<~10\degree$) when
$\ntrkone = 3$. For the second cone we required
$\dangmxtwo~<~8\degree$.  The distribution of $\dangmxtwo$ is shown in
Figure~\ref{fig:two} for the data and for the signal and background
Monte Carlo samples after the cuts on $\invmastwo$ and $\dangmxone$
have been applied.

\begin{figure}[htb]
\centerline{\epsfig{figure=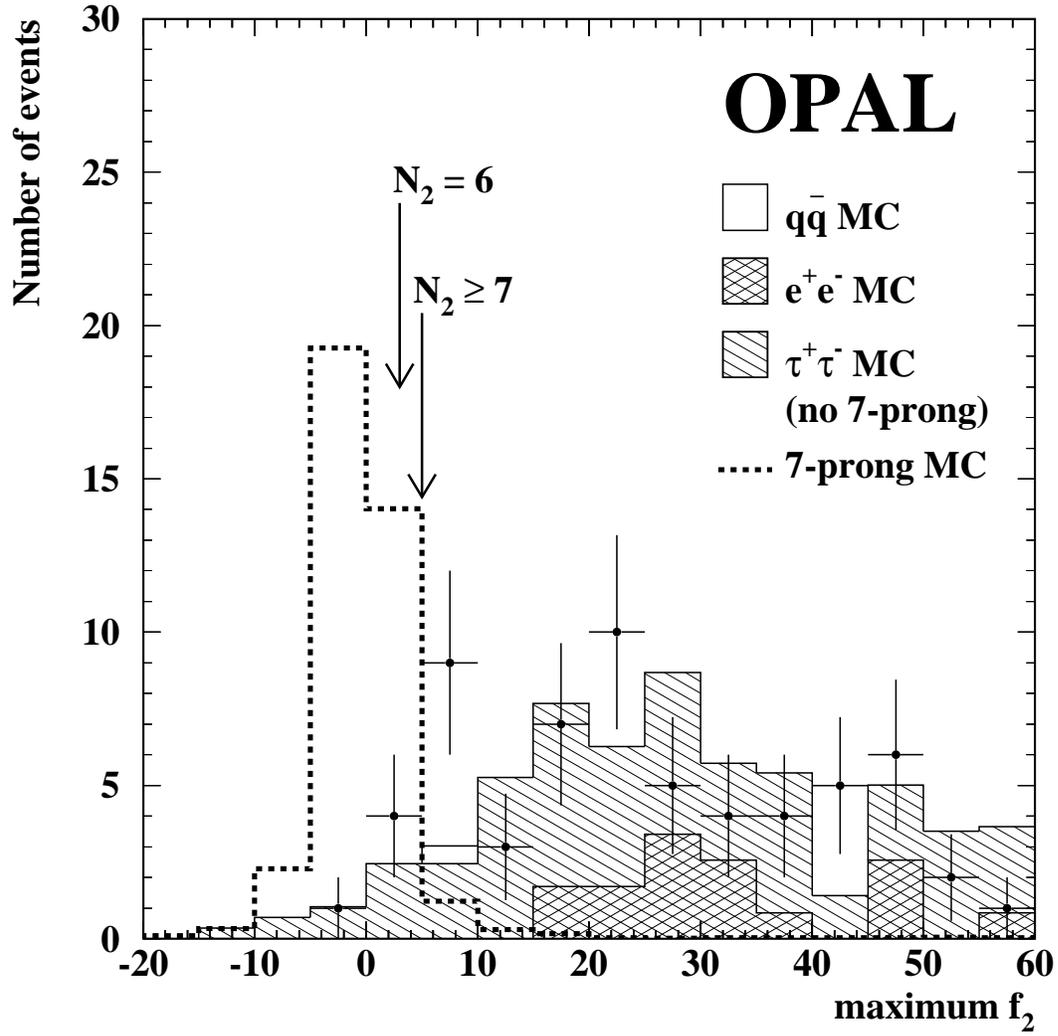,width=0.9\textwidth}}
\caption{\it Distribution of the maximum $\dedxsstwo$
before the $\dedx$ cuts. The 7-prong signal Monte Carlo simulation was
normalised assuming a branching ratio equal to the upper limit quoted
in Ref.~[4]. The other histograms show the contributions from the
various background Monte Carlo samples. The points are the data. The
arrows indicate the position of the selection cuts.}
\label{fig:three} 
\end{figure}

Jets likely to contain one or more electrons coming from $\tau$ decays
with five or fewer tracks and additional tracks from photon
conversions or $\pi^0$ Dalitz decays were rejected using the energy
loss measurement and information on the energy deposit in the
electromagnetic calorimeter.  For the tracks in the 7-prong candidate
cone we have calculated a quantity $\dedxsstwo$, which is the
logarithm of the ratio of the $\chi^2$ probability for the measured
$\dedx$ of the track to be consistent with an electron to the
probability for the measured $\dedx$ of the track to be consistent
with a pion.  If at least 20 $\dedx$ measurements along the specified
track are available, we required $\dedxsstwo$ to be less than 5.  For
the $\ntrktwo = 6$ case this cut was lowered to $\dedxsstwo~<~3$ to
reject the conversion background more effectively.  No cut was made
for tracks with less than 20 $\dedx$ measurements or for tracks where
both probabilities were smaller than 1\%.  The distributions of the
maximum $\dedxsstwo$ is shown in Figure~\ref{fig:three}.  In this
figure, the expected signal contribution is normalised assuming a
branching ratio equal to the upper limit quoted in
reference~\cite{hrs}. To reduce further the background from events
containing electrons, we applied the additional cut
$\eoptrktwo~<~0.8$, where $\econe$ is the sum of the energies of all
electromagnetic clusters inside the cone and $\pcone$ is the sum of
the momenta of all tracks associated with the cone.

After all cuts, no event is found in the data.  The expected number of
background events from all modelled sources is $0.5^{+1.0}_{-0.3}$.
The numbers of expected and observed events after each cut are shown
in Table~\ref{tab:evnumbers}.  The efficiencies for selecting a
7-prong $\tau$ event are shown in Table~\ref{tab:effi} for the signal
Monte Carlo samples described in section~\ref{evsimulation}.  A
branching ratio equal to the upper limit quoted in
reference~\cite{hrs} would result in approximately 33 $\tau$ pair
events containing a 7-prong $\tau$ decay being observed in our data
sample.

\begin{table}[htb]
\begin{center} 
\begin{tabular}{|l||r|r|r|r|c|} \hline

&\multicolumn{4}{|c|}{Number of expected events from MC}& OPAL \\ 

\raiseb{Cut}
&\multicolumn{1}{|c}{$\tautau$} 
&\multicolumn{1}{c}{$\ee$} 
&\multicolumn{1}{c}{$\qqbar$} 
&\multicolumn{1}{c|}{Sum of MC}  
&\multicolumn{1}{|c|}{data} \\ \hline \hline

Preselection,            & & & & & \\ 
$\ntrkone=1$ or 3,
& {788$\pm$13\ppo}    
& {83$\pm$7\ppo\pz} 
& {8684$\pm$79\ppo} 
& {9555$\pm$81\ppo} 
& {9593} \\
$\ntrktwo \geq 6$        & & & & & \\
\hline

$\qconeone = 1$ & & & & & \\       
$\qconetwo \leq 2$
& \raiseb{657$\pm$12\ppo}    
& \raiseb{79$\pm$7\ppo\pz} 
& \raiseb{8058$\pm$76\ppo} 
& \raiseb{8794$\pm$78\ppo} 
& \raiseb{8888} \\
\hline

%$\invmastwo(\ntrktwo=6) < 1.6~\gevcc $ & & & & & \\
%$\invmastwo(\ntrktwo\geq7) < 1.8~\gevcc $ 
invariant & & & & & \\
mass cuts 
& \raiseb{84.6$\pm$4.4}
& \raiseb{20.5$\pm$3.4}
& \raiseb{35.4$\pm$5.1}
& \raiseb{140.5$\pm$7.5}
& \raiseb{\pz131}\\
\hline

$\dangmxone < 10\degree$ & & & & & \\
$\dangmxtwo < 8\degree$
& \raiseb{50.0$\pm$3.4}
& \raiseb{18.2$\pm$3.2}
& \raiseb{ 0.6$\pm$0.6}
& \raiseb{68.8$\pm$4.7}
& \raiseb{\pzz67}\\
\hline

& & & & & \\
\raiseb{$\dedx$ cuts}
&\raiseb{3.0$\pm$0.8}     
&\raiseb{1.7$\pm$1.0}
&\raiseb{$<$ 1.1}   
&\raiseb{$4.7^{+1.7}_{-1.3}$\pz} 
&\raiseb{\pzzz4} \\ 
\hline

$\eoptrktwo$ & & & & & \\
requirement
&\raiseb{0.5$\pm$0.3}     
&\raiseb{$<$ 1.1}   
&\raiseb{$<$ 0.6}   
&\raiseb{$0.5^{+1.0}_{-0.3}$\pz} 
&\raiseb{\pzzz0} \\ 
\hline

\end{tabular} 
\caption{\it Number of expected and observed events 
  for a luminosity of 163~$\ipb$ after application of each cut.  Only
  the most important sources of background are included.  The number
  of expected background events is obtained using the Monte Carlo
  samples (MC) described in the text.  The errors shown are
  statistical only.  }
\label{tab:evnumbers} 
\end{center} 
\end{table}

\begin{table}[htb]
\begin{center} 
\begin{tabular}{|l||c|c|c|c|c|} \hline
   &\multicolumn{5}{|c|}{Efficiencies for the 7-prong selection} 
    \\ \cline{2-6}

Cut&\multicolumn{4}{|c|}{$\sevenpi$}
   &\multicolumn{1}{|c|}{$\sevenpipizero$} \\ 
%Cut&\multicolumn{4}{|c|}{$7\pi^\pm$}
%   &\multicolumn{1}{|c|}{$7\pi^\pm + \pi^0$} \\ 

   &\multicolumn{1}{|c}{isotropic}
   &\multicolumn{1}{c}{peak}
   &\multicolumn{1}{c}{high}
   &\multicolumn{1}{c|}{low}
   &\multicolumn{1}{|c|}{isotropic} \\ \hline \hline

Preselection,            & & & & & \\ 
$\ntrkone=1$ or 3,           
& 0.612$\pm$0.015    
& 0.630$\pm$0.015 
& 0.660$\pm$0.015 
& 0.627$\pm$0.015 
& 0.645$\pm$0.011 \\
$\ntrktwo \geq 6$        & & & & & \\
\hline

$\qconeone = 1$ & & & & & \\       
$\qconetwo \leq 2$
& \raiseb{0.572$\pm$0.016}    
& \raiseb{0.604$\pm$0.015} 
& \raiseb{0.618$\pm$0.015} 
& \raiseb{0.592$\pm$0.016} 
& \raiseb{0.603$\pm$0.011} \\
\hline

%$\invmastwo(\ntrktwo=6) < 1.6~\gevcc $ & & & & & \\
%$\invmastwo(\ntrktwo\geq7) < 1.8~\gevcc $ 
invariant & & & & & \\
mass cuts 
& \raiseb{0.472$\pm$0.016}
& \raiseb{0.503$\pm$0.016}
& \raiseb{0.453$\pm$0.016}
& \raiseb{0.513$\pm$0.016}
& \raiseb{0.523$\pm$0.012}\\
\hline

$\dangmxone < 10\degree$ & & & & & \\
$\dangmxtwo < 8\degree$
& \raiseb{0.447$\pm$0.016}     
& \raiseb{0.480$\pm$0.016}     
& \raiseb{0.444$\pm$0.016}     
& \raiseb{0.479$\pm$0.016}     
& \raiseb{0.493$\pm$0.012}\\
\hline

& & & & & \\
\raiseb{$\dedx$ cuts}
&\raiseb{0.419$\pm$0.016}     
&\raiseb{0.453$\pm$0.016} 
&\raiseb{0.416$\pm$0.016}
&\raiseb{0.451$\pm$0.016} 
&\raiseb{0.443$\pm$0.012} \\ 
\hline

$\eoptrktwo$ & & & & & \\
requirement
&\raiseb{0.410$\pm$0.016}     
&\raiseb{0.446$\pm$0.016} 
&\raiseb{0.404$\pm$0.016}
&\raiseb{0.445$\pm$0.016} 
&\raiseb{0.420$\pm$0.012} \\ 
\hline

\end{tabular} 
\caption{\it Efficiencies after each cut for the selection of a 7-prong 
  $\tau$ event as obtained from the different Monte Carlo simulations
  described in section~\ref{evsimulation}.  The errors shown are
  statistical only.}

\label{tab:effi} 
\end{center} 
\end{table} 

%=======================================================================
\section{Systematic errors}
%=======================================================================
\label{syserrors}

The two sources of systematic errors in this analysis are the
uncertainty in the total number of $\tau$ pair events, $\ntautau$, and
the uncertainty in the 7-prong selection efficiency.  We determined
$\ntautau$, using the total number of hadronic $\Zzero$ decays and
dividing by the ratio of the hadronic and the leptonic branching ratio
of the $\Zzero$ corrected for contributions from s-channel photon
exchange. The calculated number of $\tau$ pair events is
$\ntautau~=~\ntauevents$.  This corresponds to a systematic error of
0.5\%.

For the determination of the upper limit on the branching ratio we
chose the 7-prong Monte Carlo simulation with the isotropic phase
space distribution of the 7-pion system.  After all cuts we obtain a
selection efficiency of 0.410 for this 7-prong Monte Carlo sample.
Among all other 7-prong Monte Carlo simulations described in
section~\ref{evsimulation}, only the high phase space model leads to a
lower final selection efficiency, which is still consistent with the
selection efficiency for the isotropic phase space case. As an
estimate of the systematic uncertainty from phase space effects, we
therefore use the statistical error of the selection efficiency
obtained for the isotropic 7-prong Monte Carlo sample, corresponding
to a relative error of 3.9\%.

To take into account that additional neutral pions may be produced in
the final state, we checked differences in the selection efficiencies
for the $\sevenpi$ and the $\sevenpipizero$ decays.  Since we did not
obtain a lower selection efficiency for the $\sevenpipizero$ case, we
conclude that the branching ratio limit is relatively insensitive to
the number of additionally produced neutral pions.  The case that
kaons may be produced in the final state has been neglected because
the corresponding decay modes are expected to be highly suppressed
compared to the modes we have considered.

We studied the possible differences in the modelling of the number of
reconstructed tracks between data and the Monte Carlo simulation.
This comparison was performed using the $\dedx$ distribution for the
tracks in the 7-prong candidate cone.  In the OPAL jet chamber, for
momenta above $2~\gevc$, single tracks should have a maximum measured
$\dedx$ of 10~keV/cm (electrons) and only unresolved double tracks
should have a larger $\dedx$ value.  If at least one track with a
momentum greater than $2~\gevc$ and a $\dedx$ value greater than
15~keV/cm was found, we conclude that this is due to two tracks that
have not been resolved.  After applying the preselection and the
multiplicity cuts we calculated the fraction $\fres$ of 7-prong
candidate cones in which such a high $\dedx$ value was measured. We
found that $\fres$ is about 2\% smaller in the data than in the Monte
Carlo simulation. This 2\% discrepancy is also found in the low
invariant mass region selected with $\invmastwo < 2~\gev$.  Relaxing
the cut on $\ntrktwo$, we repeated this study for $\ntrktwo = 4$ and
$\ntrktwo = 5$ and obtained similar results. Because a higher
efficiency for the data than for the Monte Carlo simulation leads to a
systematic error which does not affect the final limit, this effect
has been neglected.

\begin{figure}[htb]
\centerline{\epsfig{figure=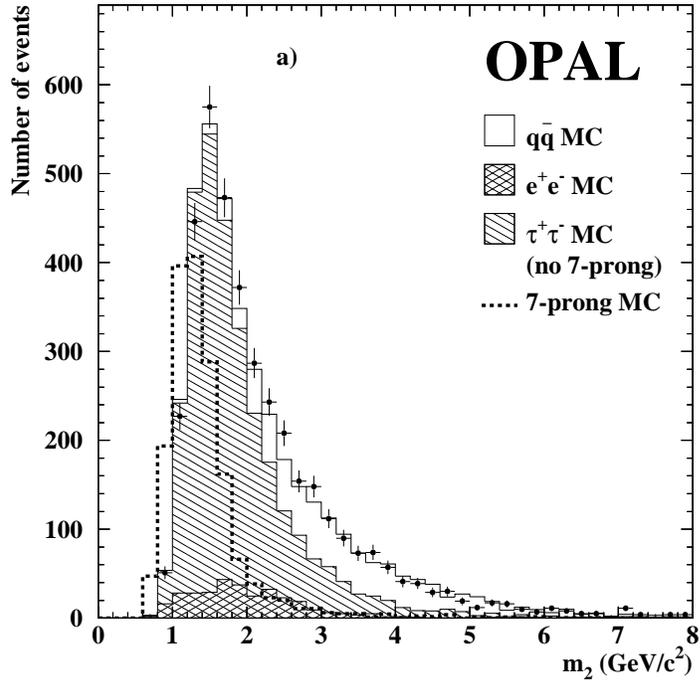,width=0.6\textwidth}}
\centerline{\epsfig{figure=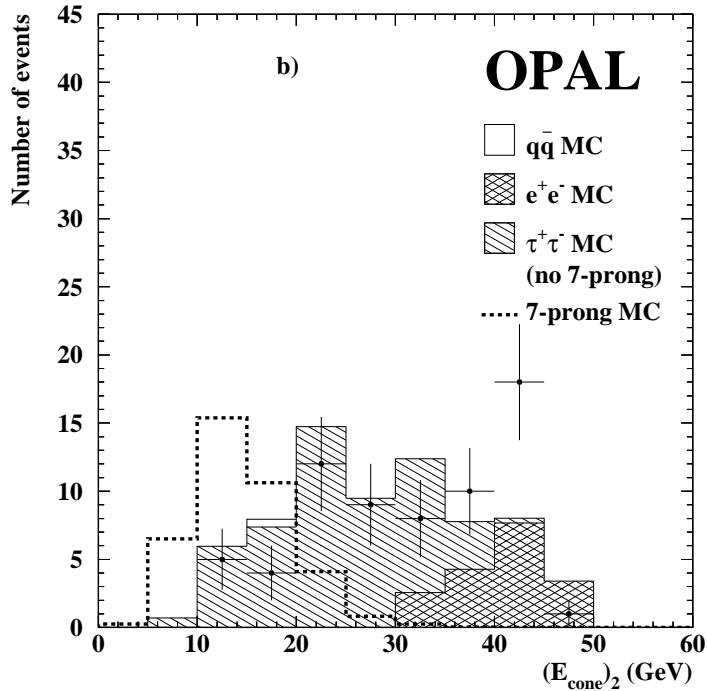,width=0.6\textwidth}}
\caption{\it Distribution of $\invmastwo$ selected by requiring 
after the preselection $\ntrkone=1$ and $5~\leq~\ntrktwo~\leq~7$ (a) and
of the total electromagnetic energy deposited in the 7-prong candidate
cone before the $\dedx$ cuts (b).  An arbitrary normalisation is used
for the 7-prong signal Monte Carlo simulation.  The other histograms
show the contributions from the various background Monte Carlo
samples. The points are the data.}
\label{fig:four} 
\end{figure}

The uncertainty arising from the cut on the total charge of the cones
was estimated by comparing the ratio of rejected events for data and
Monte Carlo simulation when the cut is applied directly after the
preselection and the multiplicity cuts, giving a relative uncertainty
of 0.7\%. For the invariant mass cut and the cut on the maximum angle
between any pair of tracks, the systematic uncertainties were
estimated by comparing data and Monte Carlo samples selected by
requiring after the preselection $\ntrkone=1$ and
$5\leq\ntrktwo\leq7$. The distribution of $\invmastwo$ after applying
these selection criteria is shown in Figure~\ref{fig:four}.  For the
two quantities, we obtain a combined relative systematic error of
1.1\%. To determine the systematic uncertainties introduced by the
cuts on $\dedxsstwo$ and $\eoptrktwo$, we have compared data and Monte
Carlo samples selected by requiring after the preselection cuts
$\ntrkone = 1$ and $\ntrktwo \geq 6$ with an additional cut on the
invariant mass in the second cone of $\invmastwo < 3~\gev$.  For these
samples, containing mostly $\tau$ pair events, a comparison of the
$\dedxsstwo$ and $\eoptrktwo$ distributions between data and the Monte
Carlo simulation leads to 7-prong efficiency uncertainties of 0.6\%
and 1\%, respectively. These results are also consistent with those
obtained in~\cite{threeprongbr}.

Adding the systematic errors in quadrature, the total relative
uncertainty of the 7-prong selection efficiency is 4.3\% and is
dominated by the uncertainty in the modelling of the 7-prong
phase space.

The reliability of the predicted background rate for $\eetoee$ events
and $\eetotautau$ events with $\tau$ decays into up to five charged
particles was effectively tested using the distribution of the total
electromagnetic energy deposited in the second cone before the $\dedx$
cuts, as shown in Figure~\ref{fig:four}. Good agreement is observed
between the data and the Monte Carlo simulation.  The number of
expected multihadronic background events has been obtained using the
JETSET prediction which has been scaled down by 10-20\% (depending on
the exact fragmentation parameters and hadronic branching ratios used
in each JETSET sample) to fit the data for events with $\ntrkone=1$ or
3 and $10 \leq \ntrktwo \leq 15$. To check the reliability of the
$\eetoqqbar$ background prediction, we have repeated the analysis
without applying the $\dedx$ and $\eoptrktwo$ cuts and reversing the
cut on $\dangmxone$.  This selects an almost pure sample of
multihadronic decays in the region where the signal is expected.
Within the limited statistics available, good agreement is found
between the data and the Monte Carlo prediction: 5 events are observed
in data and $4.8\pm1.8$ are predicted by the simulation.

%=======================================================================
\section{Results}
%=======================================================================
\label{results}

Using data collected with the OPAL detector at LEP at centre-of-mass
energies on or near the $\Zzero$ resonance, we have searched for
decays of the $\tau$ lepton into seven or more charged particles. No
candidate events were observed.  An upper limit on the branching ratio
is obtained using the calculated number of $\tau$ pair events,
$\ntautau = \ntauevents$, and the detection efficiency of ($41.0 \pm
1.8$)\% according to
\begin{equation}
\tausevenpbr < \frac{3.0}{2 \cdot \ntautau \cdot 0.410} \;\;.
\end{equation}
The total relative systematic error of
4.3\% was included in the manner described in
reference~\cite{brlimit}.  We obtain the upper limit on the branching
ratio of
\begin{equation}
\tausevenpbr < \tausevenplimit
\end{equation}
at the 95\% confidence level.  This is equivalent to an upper limit of
$1.4 \times 10^{-5}$ at the 90\% confidence level.  An improvement of
more than one order of magnitude is achieved compared to the only
previously published result~\cite{hrs}.

%=======================================================================
\section*{Acknowledgements:}
%=======================================================================
We particularly wish to thank the SL Division for the efficient operation
of the LEP accelerator and for their continuing close cooperation with
our experimental group. We thank our colleagues from CEA, DAPNIA/SPP,
CE-Saclay for their efforts over the years on the time-of-flight and trigger
systems which we continue to use.  In addition to the support staff at our own
institutions we are pleased to acknowledge the  \\
Department of Energy, USA, \\
National Science Foundation, USA, \\
Particle Physics and Astronomy Research Council, UK, \\
Natural Sciences and Engineering Research Council, Canada, \\
Israel Science Foundation, administered by the Israel
Academy of Science and Humanities, \\
Minerva Gesellschaft, \\
Benoziyo Center for High Energy Physics,\\
Japanese Ministry of Education, Science and Culture (the
Monbusho) and a grant under the Monbusho International
Science Research Program,\\
German Israeli Bi-national Science Foundation (GIF), \\
Direction des Sciences de la Mati\`ere du Commissariat \`a l'Energie
Atomique, France, \\
Bundesministerium f\"ur Bildung, Wissenschaft,
Forschung und Technologie, Germany, \\
National Research Council of Canada, \\
Hungarian Foundation for Scientific Research, OTKA T-016660,
T023793 and OTKA F-023259.\\

%=======================================================================
%       References
%=======================================================================

%=======================================================================
%       Figures
%=======================================================================

\clearpage

\clearpage

\clearpage

\clearpage
\clearpage

%%%%%%%%%%%%%%%%%%%%%%%%%%%%%%%%%%%%%%%%%%%%%%%%%%%%%%%%%%%%%%%%%%%%%%%%
\end{document}